\documentclass{ws-procs9x6}

\usepackage{tikz}
\usetikzlibrary{calc}

\begin{document}

\title{Extending Kolmogorov's axioms for a generalized probability theory on collections of contexts}

\author{Karl Svozil}

\address{Institute for Theoretical Physics, TU Wien,\\
Wiedner Hauptstrasse 8-10/136, 1040 Vienna,  Austria\\
E-mail: svozil@tuwien.ac.at\\
http://tph.tuwien.ac.at/\string~svozil}

\begin{abstract}
Kolmogorov's axioms of probability theory are extended to conditional probabilities among distinct (and sometimes intertwining) contexts. Formally, this amounts to row stochastic matrices whose entries characterize the conditional probability to find some observable (postselection) in one context, given an observable (preselection) in another context. As the respective probabilities need not (but, depending on the physical/model realization, can) be of the Born rule type, this generalizes approaches to quantum probabilities by Auff\'eves and Grangier, which in turn are inspired by Gleason's theorem.
\end{abstract}


\bodymatter

\section{Kolmogorov-type conditional probabilities among distinct contexts}

A physical system or a mathematical entity may permit not only one ``view'' on it but may allow ``many'' such views.
--
Think of a single crystal luster whose light, depending on the viewpoint, may appear very different.
Schr\"odinger~\cite[p.~15~{\&}~95]{book:1170675} quoted the Vedantic analogy of a
{\em ``many-faceted crystal which, while showing hundreds of little
pictures of what is in reality a single existent object, does
not really multiply that object. $\ldots$ A comparison used in Hinduism is of the
many almost identical images which a many-faceted
diamond makes of some one object such as the sun.''}
Another example is the coordinatization or coding and encryption of a vector with respect to different bases
--
thereby in physical terms appearing as ``coherent superpositions'' (aka linear combinations) of the respective vectors of these bases.
Still another example is the representation of an entity by isomorphic graphs.

This idea is grounded in epistemology and in issues related to the (empirical) cognition of ontology and might
appear both trivial and sophistic at first glance.
Nevertheless it may be difficult to find means or formal models exhibiting multiple contextual views of one and the same entity.
Many conceptualizations of such situations are motivated by quantum complementarity~\cite{aerts:82,aerts-69,Khrennikov2010,Dzhafarov-2017}.

A ``view'' or (used synonymously)  ``frame''~\cite{Gleason} or ``context''  will be in full generality and thus informally (glancing at heuristics from quantum mechanics and partition logic)
characterized as some domain or set of observables or properties
which are
\begin{itemize}
\item[(i)]
{\em largest} or {\em maximal} in the sense that any extension yields redundancies,
\item[(ii)]
yet at the same time in the {\em finest resolution} in the sense that the respective observables or properties are ``no composite'' of ``more elementary'' ones,
\item[(iii)]
{\em mutually exclusive} in the sense that one property or observation excludes another, different property or observation, as well as
\item[(iv)]
contains only {\em simultaneously measurable, compatible} observables or properties.
\end{itemize}

In what follows I shall develop a conceptual framework for very general probabilities on such collections of contexts.
This amounts to an extension of Kolomogorov probabilities which are defined in a single context, to a multi-context situation.
Scrutinized separately, every single context has ``legit local'' classical Kolomogorov probabilities.
In addition to those local structures and measures,
(intertwined) multi-context configurations and their
probabilities have to be ``joined'', ``woven'', ``meshed'' or ``stitched'' together to result in consistent and coherent ``global'' multi-aspect views and probabilities.

In particular,
one needs to cope with possible overlaps of contexts in common, intertwining, observables.
Because two or more contexts need not (but may) be separated from one another;
they may indeed intertwine in one or more common elements, and form complex propositional structures
serving a variety of counterfactual~\cite{specker-60,svozil-2020-c} purposes~\cite{kochen1,2015-AnalyticKS,svozil-2017-b}.

This text is organized as follows: first,
Kolmogorov's axioms or principles for probabilities are generalized to arbitrary event structures not necessarily dominated by the quantum formalism.
Then these principles will be applied to
quantum bistochasticity, as well as
partition logics which offer an abundance of alternate configurations.
Some ``exotic'' probabilities as well as possible generalizations by Cauchy's functional equations are briefly discussed.
Throughout this article, only finite contexts will be considered.

\section{Generalization of Kolmogorov's axioms to arbitrary event structures}

Suppose that, as it is assumed for classical Kolmogorov probabilities,
the elements ${\bf c}_1$  within any given single, individual finite context ${\cal C}=\{ {\bf c}_1, \ldots {\bf c}_n \}$
are mutually exclusive, compatible, and exhaustive; that is, the context contains a ``maximal'' set of mutually exclusive, compatible elements.
Kolmogorov's axioms demand that (i)  probabilities are non-negative;
(ii) additivity of mutually exclusive events or outcomes
$ P({\bf c}_i)   + P({\bf c}_j)   =  P({\bf c}_i \cup {\bf c}_j) $;
(iii) the probability of the tautology formed by the union of all elements in the context, adds up to one, that is,
$ \sum_{{\bf c}_i \in  {\cal C}}  P({\bf c}_i)= P \left(\bigcup_{{\bf c}_i \in  {\cal C}}  {\bf c}_i \right)   = 1$.

Inspired by the multi-context quantum case
discussed later the following generalization to two- or, by induction, to a multi-context configuration is suggested:
Suppose two arbitrary contexts
${\cal C}_1=\{ {\bf e}_1, \ldots {\bf e}_n \}$
and
${\cal C}_2=\{ {\bf f}_1, \ldots {\bf f}_m \}$.
The conditional probabilities
$P( {\bf f}_j \vert {\bf e}_i )$, with $1 \le j \le m$ and $1 \le i \le n$, which
alternatively can be considered as
either measuring the Bayesian degree of reasonable expectation representing
a state of knowledge or as quantification of a personal belief~\cite{Uffink2011-UFFSPS}
or the frequency of occurrence
of ``${\bf f}_j$ given ${\bf e}_i$'',
can be arranged into a $(n \times m)$-matrix whose entries are $P( {\bf f}_j \vert {\bf e}_i )$, that is,
\begin{equation}
\begin{split}
\left[ P( {\cal C}_2 \vert {\cal C}_1 )\right] =
\left[ P( \{{\bf f}_1, \ldots {\bf f}_m  \} \vert \{ {\bf e}_1, \ldots {\bf e}_n \} ) \right]
\equiv
\begin{bmatrix}
{P({\bf f}_1 \vert  {\bf e}_1)}  & \cdots & {P({\bf f}_m \vert  {\bf e}_1)}      \\
 \cdots &  \cdots &  \cdots \\
{P({\bf f}_1 \vert  {\bf e}_n)}  & \cdots &   {P({\bf f}_m \vert  {\bf e}_n)}
\end{bmatrix}
.
\end{split}
\label{2020-k-gka}
\end{equation}

Assume as axiom the following criterion:
the conditional probabilities of the elements of the second context with respect to
an arbitrary element ${\bf e}_k\in  {\cal C}_1$ of the first context ${\cal C}_1$
are non-negative, additive, and, if this sum is extended over the entire second context ${\cal C}_2$,
adds up to one:
\begin{equation}
\begin{split}
P({\bf f}_i \vert  {\bf e}_k) + P({\bf f}_j \vert  {\bf e}_k) = P[({\bf f}_i \cup {\bf f}_j) \vert  {\bf e}_k]\\
\sum_{{\bf f}_i \in  {\cal C}_2}  P({\bf f}_i \vert  {\bf e}_k) = P\left[\left(\bigcup_{{\bf f}_i \in  {\cal C}_2}  {\bf f}_i \right) \vert  {\bf e}_k\right]  = 1
.
\end{split}
\label{2020-k-gka12}
\end{equation}
That is,  the row sum taken within every single row of $\left[ P( {\cal C}_2 \vert {\cal C}_1 )\right]$ adds up to one.

This presents a generalization of Kolmogorov's axioms, as it allows cases in which both contexts do not coincide.
It just reduces to the classical axioms for single contexts if, instead of a single element  ${\bf e}_k\in  {\cal C}_1$
of the first context  ${\cal C}_1$, the union of elements of this entire context ${\cal C}_1$ -- and thus the tautology
$\bigcup_{{\bf e}_i \in  {\cal C}_1} {\bf e}_i$  -- is inserted into~(\ref{2020-k-gka12}).

We shall mostly be concerned with cases for which $n=m$;
that is, the associated matrix is a row (aka right) stochastic (square) matrix.
Formally,
such a matrix $\textsf{\textbf{A}}$ has nonnegative entries $a_{ij} \ge 0$ for $i,j =1,\ldots , n$ whose row sums
add up to one: $\sum_{j=1}^n a_{ij}=1$ for $i =1,\ldots , n$.
If, in addition to the row sums, also the column sums add up to one --
that is, if $\sum_{i=1}^n a_{ij}=1$ for $j =1,\ldots , n$ --
then the matrix is called doubly stochastic.
If $\textsf{\textbf{J}}$
is a $(n\times n)$--matrix  whose entries are $1$, then
a $(n\times n)$--matrix   $\textsf{\textbf{A}}$
is row stochastic if
$\textsf{\textbf{A}} \textsf{\textbf{J}} =  \textsf{\textbf{J}}$.

It is instructive to ponder why intuitively those conditional probabilities
should be arranged in right- but not in bistochastic matrices.
Suppose a (physical or another model) system is in a state characterized by some  element ${\bf e}_j\in  {\cal C}_1$
of the first context ${\cal C}_1$.
Then, if one takes the (union of elements of the) entire other context ${\cal C}_2$
-- thereby exhausting all possible outcomes of the second ``view'' --
the conditional probability for this system
to be in {\em any} element of ${\cal C}_2$ given ${\bf e}_j\in  {\cal C}_1$ should add up to one
because this includes all that can be (aka happen or exist) with respect to the second ``view''.
Indeed, if this conditional probability would not add up to one, say if it adds up to something strictly smaller or larger than one,
then either some elements would be missing in, or be ``external'' to, the context ${\cal C}_2$, which cannot occur since by
assumption contexts are ``maximal''.

On the other hand, if a particular element ${\bf f}_i\in  {\cal C}_f$ of the second context ${\cal C}_2$ remains fixed
and the column sum
$\sum_{{\bf e}_j \in  {\cal C}_2}  P({\bf f}_i \vert  {\bf e}_j) $
extends over all ${\bf e}_j \in  {\cal C}_2$
then there is no convincing reason why this column sum should add up to one.
Indeed, as will be argued later, while quantum mechanics results in bistochastic matrices, generalized urn models resulting in  partitions of (hidden) variables
that will not induce bistochasticity.

\section{Cauchy's functional equation encoding additivity}

One way of looking at generalized global probabilities from ``stitching'' local classical Kolmogorov probabilities is
to maintain the essence of the axioms
--
namely positivity, probability one (aka certainty) for tautologies, and, in particular, additivity.
Additivity requires that, for mutually exclusive compatible events
${\bf c}_i$
and
${\bf c}_j$ within a given context,
their probabilities can be expressed in terms of Cauchy-type functional equation
$P({\bf c}_i)   + P({\bf c}_j)   =  P({\bf c}_i \cup {\bf c}_j)$.
With ``reasonable'' side assumptions, this amounts to the linearity of probabilities in the argument~\cite{Aczel-1966,Reem2017}.

For operators in Hilbert spaces of dimensions higher than two -- and, in particular, for linear operators $\textsf{\textbf{A}}$ and $\textsf{\textbf{B}}$ with an operator norm
$\vert \textsf{\textbf{A}} \vert = + \sqrt{ \langle \textsf{\textbf{A}} \vert  \textsf{\textbf{A}} \rangle }$ based on the Hilbert-Schmidt inner product
$\langle \textsf{\textbf{A}} \vert \textsf{\textbf{B}} \rangle = \text{Trace} \left( \textsf{\textbf{A}}^\ast \textsf{\textbf{B}}\right)$,
where $\textsf{\textbf{A}}^\ast$ stands for the adjoint of $\textsf{\textbf{A}}$ --
Cauchy's functional equation can be related to Gleason-type theorems~\cite{Busch-2003,caves-fuchs-2004,Granstrom-mt,wright-Victoria,Wright_2019,Wright2019}.

The general case may involve other, hitherto unknown, arguments besides scalars and entities related to vector (or Hilbert) spaces.
The discussion will not be extended to potential inputs and sources for generalized probabilities as the main interest is in developing a generalizing probability theory in the multi-context setting,
but clearly these questions remain pertinent.

\section{Examples of application of the generalized Kolmogorov axioms}

\subsection{Quantum bistochasticity}

The multi-context quantum case has been studied in great detail with  emphasis on motivating and deriving the
Born rule~\cite{Auffeves-Grangier-2017,Auffeves-Grangier-2018} from elementary foundations.
Recall that a context has been defined as the ``largest'' or ``maximal'' domain of both   mutually exclusive  as well as   simultaneously measurable,
compatible observables.
In quantum mechanics ``simultaneously measurability'' transforms into {\em compatibility} and {\em commutativity}; that is, such observables are not complementary
and can be jointly measured without restrictions.
``Mutual exclusivity'' is defind in terms of {\em orthogonality} of the respective observables.
The spectral theorem asserts mutual orthogonality of unit eigenvectors $\vert {\bf e}_i \rangle$
and the associated orthogonal projection operators $\textsf{\textbf{E}}_i$ formed
by the dyadic product  $\textsf{\textbf{E}}_i= \vert {\bf e}_i \rangle \langle {\bf e}_i \vert$.
A context can be equivalently represented by
(i) an orthonormal basis,
(ii) the respective one-dimensional orthogonal projection operators associated with the basis elements,
or
(iii) a single maximal operator (aka maximal observable) whose spectral sum is non-degenerate~\cite{halmos-vs,kochen1}.

An essential assumption entering Gleason's derivation~\cite{Gleason}
of the Born rule for quantum probabilities is the validity of classical probability theory
whenever the respective observables are compatible.
Formally, this amounts to the validity of Kolmogorov probability theory for mutually commuting observables;
and in particular, to the assumption of Kolmogorov's axioms within contexts.

Already Gleason pointed out~\cite{Gleason} that it is quite straightforward to find an {\em ad hoc}
probability satisfying this aforementioned assumption, which is based on the Pythagorean property:
suppose (i) a quantized system is in a pure state $\vert \psi \rangle$ formalized by some unit vector,
and (ii) some ``measurement frame'' formalized by an orthonormal basis
${\cal C}=\{ \vert {\bf e}_1 \rangle , \ldots , \vert {\bf e}_n \rangle  \}$.
Then the probabilities of outcomes of observable propositions associated with the orthogonal projection operators
formed by the dyadic products $\vert {\bf e}_i \rangle \langle {\bf e}_i \vert$ of the
vectors of the orthonormal basis
can be obtained by taking the absolute square of
the length of those projections of $\vert \psi \rangle$ onto $\vert {\bf e}_i \rangle$ along the remaining basis vectors,
which amounts to taking the scalar products
$\vert \langle \psi \vert {\bf e}_i \rangle \vert^2$.
Since the vector associated with the pure state as well as all the vectors in the orthonormal system are of length one,
and since these latter vectors (of the orthonormal system) are mutually orthogonal,
the sum
$\sum_{i=1}^n \vert \langle \psi \vert {\bf e}_i \rangle \vert^2$
of all these terms, taken over all the basis elements, needs to add up to one.
The respective absolute squares are bounded between zero and one.
In effect, the orthonormal basis ``grants a view'' of the pure quantum state.
The absolute square can be rewritten in terms of a trace (over some arbitrary orthonormal basis)
into the standard form known as the Born rule of quantum probabilities:
$
\vert \langle \psi \vert {\bf e}_i \rangle \vert^2
=
\langle \psi \vert {\bf e}_i \rangle \langle {\bf e}_i \vert \psi \rangle
=
\langle \psi \vert {\bf e}_i \rangle \langle {\bf e}_i \vert \mathbb{I}_n \psi \rangle
=
\sum_{j=1}^n \langle \psi \vert {\bf e}_i \rangle \langle {\bf e}_i \vert {\bf g}_j \rangle \langle {\bf g}_j \vert \psi \rangle
=
\sum_{j=1}^n \langle {\bf g}_j \vert \underbrace{\psi \rangle\langle \psi}_{=\textsf{\textbf{E}}_\psi} \vert \underbrace{ {\bf e}_i \rangle \langle {\bf e}_i }_{=\textsf{\textbf{E}}_i} \vert {\bf g}_j \rangle
=
\text{Trace}(\textsf{\textbf{E}}_\psi \textsf{\textbf{E}}_i)
$,
where
$\textsf{\textbf{E}}_\psi$ and $\textsf{\textbf{E}}_i$
are the orthogonal projection operators representing the state
$\vert \psi \rangle$ and the (unit) vectors of the orthonormal basis $\vert {\bf e}_i\rangle$, respectively,
and
${\cal C}'=\{ \vert {\bf g}_1 \rangle , \ldots , \vert {\bf g}_n \rangle  \}$
is an arbitrary orthonormal basis, so that a resolution of the identity is
$ \mathbb{I}_n = \sum_{j=1}^n \vert {\bf g}_j \rangle \langle {\bf g}_j \vert$.

It is also well known that, at least from a formal perspective, unit vectors in quantum mechanics serve a dual role:
On the one hand, they represent pure states.
On the other hand, by the associated one-dimensional orthogonal projection operator, they represent an observable:
the proposition that the system is in such a pure state~\cite{v-neumann-49,birkhoff-36}.
Suppose now that we exploit this dual role by {\em expanding}  the pure prepared state into a full orthonormal basis,
of which its vector must be an element.
(For dimensions greater than two such an expansion will not be unique as there is a continuous infinity of ways to achieve this.)
Once the latter basis is fixed it can be used to obtain a ``view'' on the former (measurement) basis;
and a completely symmetric situation/configuration is attained.
We might even go so far as to say that which basis is associated with the ``observed object'' and with the ``measurement apparatus,''
respectively,
is purely a matter of convention and subjective perspective.

Therefore, as has been pointed out earlier, an orthogonal projection operator
serves a dual role:
on the one hand it is a formalization of a dichotomic observable -- more precisely, an elementary yes-no proposition
$\textsf{\textbf{E}} =  \vert {\bf x} \rangle \langle   {\bf x}   \vert $
associated with the claim that ``the quantized system is in state $\vert {\bf x} \rangle$.
And on the other hand it is the formal representation of a pure quantum state $\vert {\bf y} \rangle$,
equivalent to the operator $\textsf{\textbf{F}} =  \vert {\bf y} \rangle \langle   {\bf y}   \vert $.
By the Born rule the conditional probabilities are symmetric with respect to exchange of
$\vert {\bf x} \rangle$
and
$\vert {\bf y} \rangle$:
let ${\cal C}'=\{ \vert {\bf g}_1 \rangle , \ldots , \vert {\bf g}_n \rangle  \}$
be some arbitrary orthonormal basis of $\mathbb{C}^n$, then
$P( \textsf{\textbf{E}} \vert \textsf{\textbf{F}} )
=\text{Trace} \left( \textsf{\textbf{E}} \textsf{\textbf{F}}\right)
=\text{Trace} \left( \textsf{\textbf{F}} \textsf{\textbf{E}}\right)
=P( \textsf{\textbf{F}} \vert \textsf{\textbf{E}} )$; or, more explicitly,
$P( \textsf{\textbf{E}} \vert \textsf{\textbf{F}} )
=  \sum_{i=1}^n
\langle  {\bf g}_i \vert  {\bf x} \rangle
\langle   {\bf x}   \vert {\bf y} \rangle
\langle   {\bf y} \vert {\bf g}_i \rangle
=  \sum_{i=1}^n
\langle   {\bf x}   \vert {\bf y} \rangle
\langle   {\bf y} \vert \underbrace{{\bf g}_i \rangle \langle  {\bf g}_i}_{=\mathbb{I}_n} \vert  {\bf x} \rangle
= \left| \langle   {\bf x}   \vert {\bf y} \rangle \right|^2
= \left| \langle   {\bf y}   \vert {\bf x} \rangle \right|^2
=P( \textsf{\textbf{F}} \vert \textsf{\textbf{E}} )
$.
Therefore, the respective conditional probabilities form a doubly stochastic (bistochastic) square matrix.
This result is a special case of a more general result
on quadratic forms on the set of eigenvectors of normal operators~\cite{marcus-1960}.

Consider two orthonormal bases aka two contexts.
Their respective conditional probabilities can be arranged into a matrix form:
The $i$th row $j$th column component corresponds to the conditional probability
associated with the probability of occurrence of the $j$th element (observable) of the second context,
given the $i$th element (observable)  of the first context.
By taking into account that cyclically interchanging factors inside a trace does not change its value
this matrix needs to be not only row (right) stochastic but doubly stochastic (bistochastic)~\cite{Auffeves-Grangier-2017,Auffeves-Grangier-2018};
that is, the sum is taken within every single row and every single column adds up to one.

\subsection{Quasi-classical partition logics}

In what follows we shall study sets of partitions of a given set.
They have models~\cite{svozil-2001-eua} based (i) on the finite automata initial state identification problem~\cite{e-f-moore}
as well as (ii) on generalized urns~\cite{wright:pent,wright}.
Partition logics are quasi-classical and value-definite in so far as they allow a separating set of ``classical'' two-valued states~\cite[Theorem~0]{kochen1};
and yet they feature complementarity.
Many of these logics are {\it doubles} of quantum logics, such as for spin-state measurements;
and thereby their graphs also allow faithful orthogonal representations~\cite{svozil-2018-b};
and yet some of them have no quantum analog.
Therefore, they neither form a proper subset of all quantum logics nor do they contain all logical structures encountered in quantum logics
(they are neither continuous nor can they have a non-separating or nonexisting set of two-valued states).
However, partition logics overlaps significantly with quantum logics,
as they bear strong similarities with the structures arising in quantum theory.

If some (partition) logic which is a pasting~\cite{greechie:71,kalmbach-83,nav:91} of contexts
has a separating set of two-valued states~\cite[Theorem~0]{kochen1} then there is a
constructive, algorithmic~\cite{tkadlec-2states-2017} way of finding a ``canonical'' partition logic~\cite{svozil-2001-eua},
and, associated with it, all classical probabilities
on it: first, find all the two-valued states on the logic, and assign consecutive number to these states.
Then, for any atom (element of a context), find the index set of all two-valued states which are~1
on this atom.
Associate with each one, say, the $i$th, of the two valued states a nonnegative weight $i \rightarrow \lambda_i$,
and require that the (convex) sum of these weights $\sum_i \lambda_i=1$ is $1$.
Since all two-valued states are included, the Kolmogorov axioms guarantee that the sum of measures/weights
within each of the contexts in the logic exactly adds up to one.

It will be argued that in this case, and unlike for quantum conditional probabilities,
the conditional probabilities, in general, do not form a bistochastic matrix.

\subsubsection{Two non-intertwining two-atomic contexts}

In the Babylonian spirit~\cite[p.~172]{neugeb}
consider some anecdotal examples which have quantum doubles.
The first one will be analogous to a spin-$\frac{1}{2}$ state measurement.

\begin{figure}[htpb]
\begin{center}
\begin{tabular}{cc}
\begin{tikzpicture}  [scale=0.3]

\newdimen\ms
\ms=0.05cm

\tikzstyle{every path}=[line width=1.5pt]

\tikzstyle{c3}=[circle,inner sep={\ms/8},minimum size=4.5*\ms]
\tikzstyle{c2}=[circle,inner sep={\ms/8},minimum size=3*\ms]
\tikzstyle{c1}=[circle,inner sep={\ms/8},minimum size=1.5*\ms]

\path
  (0,0) coordinate(1)
  (6,0 ) coordinate(2)
  (0,6 ) coordinate(3)
  (6,6  ) coordinate(4)
;


\draw [color=orange] (1) -- (2);
\draw [color=blue] (3) -- (4);

%
%
\draw (1) coordinate[c3,fill=orange,label={above:\footnotesize $\{1,3\}$}];   %
\draw (2) coordinate[c3,fill=orange,label={above:\footnotesize $\{2,4\}$}];    %
\draw (3) coordinate[c3,fill=blue,label={above:\footnotesize $\{1,2\}$}]; %
\draw (4) coordinate[c3,fill=blue,label={above:\footnotesize $\{3,4\}$}];  %

\end{tikzpicture}
&
\begin{tikzpicture}  [scale=0.3]

\newdimen\ms
\ms=0.05cm

\tikzstyle{every path}=[line width=1.5pt]

\tikzstyle{c3}=[circle,inner sep={\ms/8},minimum size=4.5*\ms]
\tikzstyle{c2}=[circle,inner sep={\ms/8},minimum size=3*\ms]
\tikzstyle{c1}=[circle,inner sep={\ms/8},minimum size=1.5*\ms]

\path
  (0,0) coordinate(1)
  (6,0 ) coordinate(2)
  (0,6 ) coordinate(3)
  (6,6  ) coordinate(4)
;


\draw [color=orange] (1) -- (2);
\draw [color=blue] (3) -- (4);

%
%
\draw (1) coordinate[c3,fill=orange,label={above:\footnotesize $\frac{1}{\sqrt{2}}\begin{pmatrix}1,1\end{pmatrix}^\intercal$}];   %
\draw (2) coordinate[c3,fill=orange,label={above:\footnotesize $\frac{1}{\sqrt{2}}\begin{pmatrix}1,-1\end{pmatrix}^\intercal$}];    %
\draw (3) coordinate[c3,fill=blue,label={above:\footnotesize $\begin{pmatrix}1,0\end{pmatrix}^\intercal$}]; %
\draw (4) coordinate[c3,fill=blue,label={above:\footnotesize $\begin{pmatrix}0,1\end{pmatrix}^\intercal$}];  %

\end{tikzpicture}
\\
(a)&(b)
\end{tabular}
\end{center}
\caption{Greechie orthogonality diagram of a logic consisting of two nonintertwining contexts.
(a)
The associated (quasi)classical partition logic representations
obtained by an inverse construction using
all two-valued measures thereon~\cite{svozil-2001-eua};
(b) a faithful orthogonal representation~\cite{lovasz-79}
rendering a quantum {\it double}.
}
\label{2019-k-f1}
\end{figure}
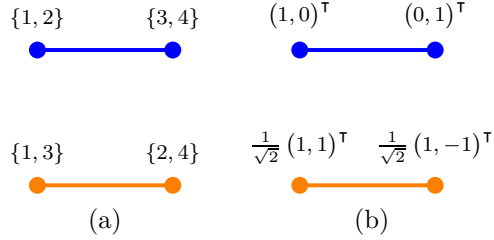

The   logic in Fig.~\ref{2019-k-f1} enumerates the labels of the atoms (aka elementary propositions)
according to the ``inverse construction'' -- based on all four two-valued states on the logic -- mentioned earlier,
using all two-valued measures thereon~\cite{svozil-2001-eua}.
With the identifications
${\bf e}_1 \equiv \{1,2\}$,
${\bf e}_2 \equiv \{3,4\}$,
${\bf f}_1 \equiv \{1,3\}$, and
${\bf f}_2 \equiv \{2,4\}$
we obtain all classical probabilities
by identifying $i \rightarrow \lambda_i > 0$.
The respective conditional probabilities are
\begin{equation}
\begin{split}
\left[ P( {\cal C}_2 \vert {\cal C}_1 )\right] =
\left[ P( \{ {\bf f}_1, {\bf f}_2 \} \vert \{ {\bf e}_1,{\bf e}_2 ) \right]
\equiv
\begin{bmatrix}
{P({\bf f}_1 \vert  {\bf e}_1)}  &  {P({\bf f}_2 \vert  {\bf e}_1)}      \\
{P({\bf f}_1 \vert  {\bf e}_2)}  &  {P({\bf f}_2 \vert  {\bf e}_2)}
\end{bmatrix}
\\
=
\begin{bmatrix}
\frac{P({\bf f}_1 \cap {\bf e}_1)}{P({\bf e}_1)}  &  \frac{P({\bf f}_2 \cap {\bf e}_1)}{P({\bf e}_1)}      \\
\frac{P({\bf f}_1 \cap {\bf e}_2)}{P({\bf e}_2)}  &  \frac{P({\bf f}_2 \cap {\bf e}_2)}{P({\bf e}_2)}
\end{bmatrix}
=
\begin{bmatrix}
\frac{P( \{1,3\}  \cap  \{1,2\} )}{P( \{1,2\} )}  &  \frac{P( \{2,4\}  \cap  \{1,2\} )}{P( \{1,2\} )}     \\
\frac{P( \{1,3\}  \cap  \{3,4\} )}{P( \{3,4\} )}  &  \frac{P( \{2,4\}  \cap  \{3,4\} )}{P( \{3,4\} )}
\end{bmatrix}
\\
=
\begin{bmatrix}
\frac{P( \{1\} )}{P( \{1,2\} )}  &  \frac{P( \{2\} )}{P( \{1,2\} )}   \\
\frac{P( \{3\} )}{P( \{3,4\} )}  &  \frac{P( \{4\} )}{P( \{3,4\} )}
\end{bmatrix}
=
\begin{bmatrix}
\frac{ \lambda_1 }{ \lambda_1+\lambda_2 }  &  \frac{ \lambda_2 }{ \lambda_1+\lambda_2 }     \\
\frac{ \lambda_3 }{ \lambda_3+\lambda_4 }  &  \frac{ \lambda_4 }{ \lambda_3+\lambda_4 }
\end{bmatrix},
\end{split}
\end{equation}
as well as
\begin{equation}
\begin{split}
\left[ P( {\cal C}_1 \vert {\cal C}_2 )\right] =
\left[ P( \{ {\bf e}_1, {\bf e}_2 \} \vert \{ {\bf f}_1,{\bf f}_2  \} ) \right]
\\
\equiv
\begin{bmatrix}
\frac{P( \{1\} )}{P( \{1,3\} )}  &  \frac{P( \{3\} )}{P( \{1,3\} )}      \\
\frac{P( \{2\} )}{P( \{2,4\} )}  &  \frac{P( \{4\} )}{P( \{2,4\} )}
\end{bmatrix}
=
\begin{bmatrix}
\frac{ \lambda_1 }{ \lambda_1+\lambda_3 }  &  \frac{ \lambda_3 }{ \lambda_1+\lambda_3 }       \\
\frac{ \lambda_2 }{ \lambda_2+\lambda_4 }  &  \frac{ \lambda_4 }{ \lambda_2+\lambda_4 }
\end{bmatrix}
.
\end{split}
\end{equation}

\subsubsection{Two intertwining three-atomic contexts}

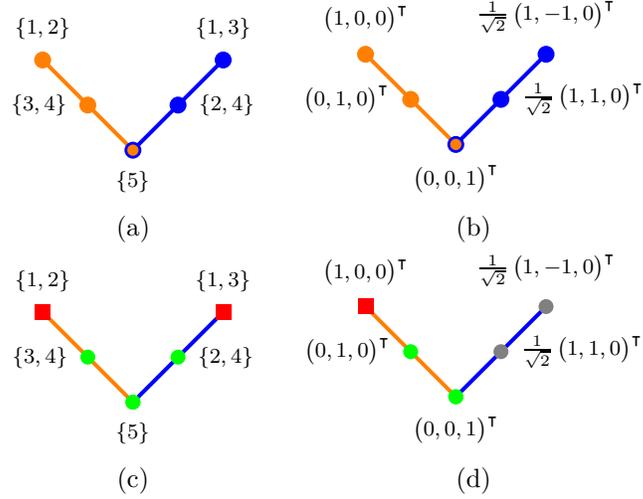
\begin{figure}[htpb]
\begin{center}
\begin{tabular}{cc}
\begin{tikzpicture}  [scale=0.2]

\newdimen\ms
\ms=0.05cm

\tikzstyle{every path}=[line width=1.5pt]

\tikzstyle{c3}=[circle,inner sep={\ms/8},minimum size=4.5*\ms]
\tikzstyle{c2}=[circle,inner sep={\ms/8},minimum size=3*\ms]
\tikzstyle{c1}=[circle,inner sep={\ms/8},minimum size=1.5*\ms]

\newdimen\R
\R=6cm     



\path
  (0,6 ) coordinate(1)
  (3,3    ) coordinate(2)
  (6,0 ) coordinate(3)
  (9,3) coordinate(4)
  (12,6  ) coordinate(5)
;


\draw [color=orange] (1) -- (2) -- (3);
\draw [color=blue] (3) -- (4) -- (5);

%
%
\draw (1) coordinate[c3,fill=orange,label={above:\footnotesize $\{1,2\}$}];   %
\draw (2) coordinate[c3,fill=orange,label={left:\footnotesize $\{3,4\}$}];    %
\draw (3) coordinate[c3,fill=blue,label={below:\footnotesize $\{5\}$}]; %
\draw (3) coordinate[c2,fill=orange];  %
\draw (4) coordinate[c3,fill=blue,label={right:\footnotesize $\{2,4\}$}];  %
\draw (5) coordinate[c3,fill=blue,label={above:\footnotesize $\{1,3\}$}];  %

\end{tikzpicture}
&
\begin{tikzpicture}  [scale=0.2]

\newdimen\ms
\ms=0.05cm

\tikzstyle{every path}=[line width=1.5pt]

\tikzstyle{c3}=[circle,inner sep={\ms/8},minimum size=4.5*\ms]
\tikzstyle{c2}=[circle,inner sep={\ms/8},minimum size=3*\ms]
\tikzstyle{c1}=[circle,inner sep={\ms/8},minimum size=1.5*\ms]

\newdimen\R
\R=6cm     



\path
  (0,6 ) coordinate(1)
  (3,3    ) coordinate(2)
  (6,0 ) coordinate(3)
  (9,3) coordinate(4)
  (12,6  ) coordinate(5)
;


\draw [color=orange] (1) -- (2) -- (3);
\draw [color=blue] (3) -- (4) -- (5);

%
%
\draw (1) coordinate[c3,fill=orange,label={above:\footnotesize $\begin{pmatrix}1,0,0\end{pmatrix}^\intercal$}];   %
\draw (2) coordinate[c3,fill=orange,label={left:\footnotesize  $\begin{pmatrix}0,1,0\end{pmatrix}^\intercal$}];    %
\draw (3) coordinate[c3,fill=blue,label={below:\footnotesize    $\begin{pmatrix}0,0,1\end{pmatrix}^\intercal$}]; %
\draw (3) coordinate[c2,fill=orange];  %
\draw (4) coordinate[c3,fill=blue,label={right:\footnotesize    $\frac{1}{\sqrt{2}}\begin{pmatrix}1,1,0\end{pmatrix}^\intercal$}];  %
\draw (5) coordinate[c3,fill=blue,label={above:\footnotesize    $\frac{1}{\sqrt{2}}\begin{pmatrix}1,-1,0\end{pmatrix}^\intercal$}];  %

\end{tikzpicture}
\\
(a)&(b)
\\
\begin{tikzpicture}  [scale=0.2]

\newdimen\ms
\ms=0.05cm

\tikzstyle{every path}=[line width=1.5pt]

\tikzstyle{s1}=[color=red,inner sep=2,rectangle,minimum size=6]
\tikzstyle{c2}=[circle,inner sep=2,minimum size=4]

\newdimen\R
\R=6cm     



\path
  (0,6 ) coordinate(1)
  (3,3    ) coordinate(2)
  (6,0 ) coordinate(3)
  (9,3) coordinate(4)
  (12,6  ) coordinate(5)
;


\draw [color=orange] (1) -- (2) -- (3);
\draw [color=blue] (3) -- (4) -- (5);

%
%
\draw (1) coordinate[s1,fill=red,label={above:\footnotesize $\{1,2\}$}];   %
\draw (2) coordinate[c2,fill=green,label={left:\footnotesize $\{3,4\}$}];    %
\draw (3) coordinate[c2,fill=green,label={below:\footnotesize $\{5\}$}]; %
\draw (3) coordinate[c2,fill=green];  %
\draw (4) coordinate[c2,fill=green,label={right:\footnotesize $\{2,4\}$}];  %
\draw (5) coordinate[s1,fill=red,label={above:\footnotesize $\{1,3\}$}];  %

\end{tikzpicture}
&
\begin{tikzpicture}  [scale=0.2]

\newdimen\ms
\ms=0.05cm

\tikzstyle{every path}=[line width=1.5pt]

\tikzstyle{s1}=[color=red,inner sep=2,rectangle,minimum size=6]
\tikzstyle{c2}=[circle,inner sep=2,minimum size=4]

\newdimen\R
\R=6cm     



\path
  (0,6 ) coordinate(1)
  (3,3    ) coordinate(2)
  (6,0 ) coordinate(3)
  (9,3) coordinate(4)
  (12,6  ) coordinate(5)
;


\draw [color=orange] (1) -- (2) -- (3);
\draw [color=blue] (3) -- (4) -- (5);

%
%
\draw (1) coordinate[s1,fill=red,label={above:\footnotesize   $\begin{pmatrix}1,0,0\end{pmatrix}^\intercal$}];   %
\draw (2) coordinate[c2,fill=green,label={left:\footnotesize  $\begin{pmatrix}0,1,0\end{pmatrix}^\intercal$}];    %
\draw (3) coordinate[c2,fill=green,label={below:\footnotesize $\begin{pmatrix}0,0,1\end{pmatrix}^\intercal$}]; %
\draw (3) coordinate[c2,fill=green];  %
\draw (4) coordinate[c2,fill=gray,label={right:\footnotesize  $\frac{1}{\sqrt{2}}\begin{pmatrix}1,1,0\end{pmatrix}^\intercal$}];  %
\draw (5) coordinate[c2,fill=gray,label={above:\footnotesize  $\frac{1}{\sqrt{2}}\begin{pmatrix}1,-1,0\end{pmatrix}^\intercal$}];  %

\end{tikzpicture}
\\
(c)&(d)
\end{tabular}
\end{center}
\caption{Greechie orthogonality diagram of  the $L_{12}$ ``firefly'' logic.
(a) The associated (quasi)classical partition logic representation
obtained through in inverse construction using all two-valued measures thereon~\cite{svozil-2001-eua};
(b) a faithful orthogonal representation~\cite{lovasz-79}
rendering a quantum {\it double};
(c) ``classical'' two-valued measure number $1$;
(d) a pure quantum state prepared as $\begin{pmatrix}1,0,0\end{pmatrix}^\intercal$.
A red square and gray and green circles indicate value assignments $1$, $\frac{1}{2}$ and $0$, respectively.
}
\label{2019-k-f-ffl}
\end{figure}

The  $L_{12}$ ``firefly'' logic depicted in Fig.~\ref{2019-k-f-ffl} labels the atoms (aka elementary propositions)
obtained by an ``inverse construction''
using all five two-valued measures thereon~\cite{svozil-2001-eua,svozil-2016-s}.
By design, it will be very similar to the earlier logic with four atoms.
With the identifications
${\bf e}_1 \equiv \{1,2\}$,
${\bf e}_2 \equiv \{3,4\}$,
${\bf e}_3 = {\bf f}_3 \equiv \{5\}$,
${\bf f}_1 \equiv \{1,3\}$, and
${\bf f}_2 \equiv \{2,4\}$
we obtain all classical probabilities
by identifying $i \rightarrow \lambda_i > 0$.
The respective conditional probabilities are
\begin{equation}
\begin{split}
\left[ P( {\cal C}_2 \vert {\cal C}_1 )\right] =
\left[ P( \{ {\bf f}_1, {\bf f}_2, {\bf f}_3 \} \vert \{ {\bf e}_1,{\bf e}_2,{\bf e}_3 \} ) \right]
\\
\equiv
\begin{bmatrix}
\frac{P( \{1\} )}{P( \{1,2\} )}  &  \frac{P( \{2\} )}{P( \{1,2\} )}   & \frac{P( \emptyset )}{P( \{1,2\} )}    \\
\frac{P( \{3\} )}{P( \{3,4\} )}  &  \frac{P( \{4\} )}{P( \{3,4\} )}   & \frac{P( \emptyset  )}{P( \{3,4\} )}    \\
\frac{P( \emptyset  )}{P( \{5\} )}  &  \frac{P( \emptyset  )}{P( \{5\} )}   & \frac{P( \{5\} )}{P( \{5\} )}
\end{bmatrix}
=
\begin{bmatrix}
\frac{ \lambda_1 }{ \lambda_1+\lambda_2 }  &  \frac{ \lambda_2 }{ \lambda_1+\lambda_2 }   & 0    \\
\frac{ \lambda_3 }{ \lambda_3+\lambda_4 }  &  \frac{ \lambda_4 }{ \lambda_3+\lambda_4 }   & 0    \\
0  &  0   & 1
\end{bmatrix},
\end{split}
\end{equation}
as well as
\begin{equation}
\begin{split}
\left[ P( {\cal C}_1 \vert {\cal C}_2 )\right] =
\left[ P( \{ {\bf e}_1, {\bf e}_2, {\bf e}_3 \} \vert \{ {\bf f}_1,{\bf f}_2,{\bf f}_3 \} ) \right]
\\
\equiv
\begin{bmatrix}
\frac{P( \{1\} )}{P( \{1,3\} )}  &  \frac{P( \{3\} )}{P( \{1,3\} )}   & \frac{P( \emptyset )}{P( \{1,3\} )}    \\
\frac{P( \{2\} )}{P( \{2,4\} )}  &  \frac{P( \{4\} )}{P( \{2,4\} )}   & \frac{P( \emptyset  )}{P( \{2,4\} )}    \\
\frac{P( \emptyset  )}{P( \{5\} )}  &  \frac{P( \emptyset  )}{P( \{5\} )}   & \frac{P( \{5\} )}{P( \{5\} )}
\end{bmatrix}
=
\begin{bmatrix}
\frac{ \lambda_1 }{ \lambda_1+\lambda_3 }  &  \frac{ \lambda_3 }{ \lambda_1+\lambda_3 }   & 0    \\
\frac{ \lambda_2 }{ \lambda_2+\lambda_4 }  &  \frac{ \lambda_4 }{ \lambda_2+\lambda_4 }   & 0    \\
0  &  0   & 1
\end{bmatrix}
.
\end{split}
\label{2019-k-e-fireflyrsm}
\end{equation}

The conditional probabilities of the firefly logic,
as depicted in Fig.~\ref{2019-k-f-ffl}(a), and  enumerated in Eq.~(\ref{2019-k-e-fireflyrsm})
form a  right stochastic  matrix.
As mentioned earlier, given
any particular outcome ${\bf f}_i$ of the second context
corresponding to some respective row in the matrix~(\ref{2019-k-e-fireflyrsm}),
the row-sum of the conditional probabilities of all the conceivable mutually exclusive outcomes
of the first context
$ {\cal C}_1 = \{ {\bf e}_1,{\bf e}_2,{\bf e}_3 \}$  must be one.
However, the ``transposed'' statement is not true:
the column-sum of the conditional probabilities
of a particular element ${\bf e}_j$ with respect
to all the mutually exclusive outcomes of the second context $ {\cal C}_2 = \{ {\bf f}_1,{\bf f}_2,{\bf f}_3 \}$,
needs not be one.

Take, for example, the singular distribution case such that $\lambda_1=1$, and therefore, by positivity and convexity,
$\lambda_{i \neq 1}=0$, that is, $\lambda_2=\lambda_3=\lambda_4=\lambda_5=0$.
This configuration, depicted in Fig.~\ref{2019-k-f-ffl}(c),
results in the following, partial (undefined components are indicated by the symbol ``$\frac{0}{0}$'')
right stochastic matrix~(\ref{2019-k-e-fireflyrsm-be})
derived from~(\ref{2019-k-e-fireflyrsm}):
\begin{equation}
\begin{bmatrix}
\frac{P( \{1\} )}{P( \{1,3\} )}  &  \frac{P( \{3\} )}{P( \{1,3\} )}   & \frac{P( \emptyset )}{P( \{1,3\} )}    \\
\frac{P( \{2\} )}{P( \{2,4\} )}  &  \frac{P( \{4\} )}{P( \{2,4\} )}   & \frac{P( \emptyset  )}{P( \{2,4\} )}    \\
\frac{P( \emptyset  )}{P( \{5\} )}  &  \frac{P( \emptyset  )}{P( \{5\} )}   & \frac{P( \{5\} )}{P( \{5\} )}
\end{bmatrix}
=
\begin{bmatrix}
\frac{ \lambda_1 }{ \lambda_1+\lambda_3 }  &  \frac{ \lambda_3 }{ \lambda_1+\lambda_3 }   & 0    \\
\frac{ \lambda_2 }{ \lambda_2+\lambda_4 }  &  \frac{ \lambda_4 }{ \lambda_2+\lambda_4 }   & 0    \\
0  &  0   & 1
\end{bmatrix}
=
\begin{bmatrix}
1  &  0  & 0    \\
\frac{0}{0}  &  \frac{0}{0}   & 0    \\
0  &  0   & 1
\end{bmatrix}
.
\label{2019-k-e-fireflyrsm-be}
\end{equation}
In such a case, in terms of, say, a generalized urn model, the observable proposition
$\{2,4\}$ associated with the plaintext
{\em ``looked upon in the first color (in this case blue), the ball drawn from the urn
shows the symbols $2$ or $4$''}
will never occur; regardless of which ball type associated with the other context
$\{1,2\}$, $\{3,4\}$, or $\{5\}$ one would have (counterfactually) drawn
because the generalized urn is only loaded with balls of one type,
namely the first type, with the symbol ``$\{1,2\}$'' painted on them in the first color,
and the symbols ``$\{1,3\}$'' painted on them in the second color.
(Instead of labels indicating the elements of the partition one may choose other symbols, such as
$\{1,3\} \equiv a \equiv \{1,2\}$,
$\{2,4\} \equiv b \equiv \{3,4\}$, and
$c \equiv \{5\}$ in the respective colors~\cite{svozil-2001-eua,svozil-2008-ql}.)

Ultimately one may say that it is the {\em discontinuity} of the two-valued measures which ``prevents''
the quasiclassical conditional probabilities to be arranged in a bistochastic matrix.
A similar quantum realization could, for instance, be obtained by the three-dimensional
faithful orthogonal representation~\cite{lovasz-79}
$\{1,2\} \equiv \begin{pmatrix}1,0,0\end{pmatrix}^\intercal$,
$\{3,4\} \equiv \begin{pmatrix}0,1,0\end{pmatrix}^\intercal$,
$\{5\} \equiv \begin{pmatrix}0,0,1\end{pmatrix}^\intercal$,
$\{1,3\} \equiv (1/\sqrt{2})\begin{pmatrix}1,1,0\end{pmatrix}^\intercal$, and
$\{2,4\} \equiv (1/\sqrt{2})\begin{pmatrix}1,-1,0\end{pmatrix}^\intercal$.
Preparition (aka ``loading the quantum urn'') with state $\{1,2\} \equiv \begin{pmatrix}1,0,0\end{pmatrix}^\intercal$,
as depicted in Fig.~\ref{2019-k-f-ffl}(d),
yields the quantum bistochastic matrix
\begin{equation}
\begin{split}
\left[ P \left( \left\{ \begin{pmatrix}1 \\ 0 \\ 0\end{pmatrix} , \begin{pmatrix}0 \\ 1 \\ 0\end{pmatrix} ,\begin{pmatrix}0 \\ 0 \\ 1\end{pmatrix}  \right\}
\middle\vert
\left\{\frac{1}{\sqrt{2}}\begin{pmatrix}1 \\ 1 \\ 0\end{pmatrix} ,  \frac{1}{\sqrt{2}}\begin{pmatrix}1 \\ -1 \\ 0\end{pmatrix}  ,\begin{pmatrix}0 \\ 0 \\ 1\end{pmatrix}  \right\} \right) \right]
\\
=
\begin{bmatrix}
\frac{1}{ {2}}  &  \frac{1}{ {2}}   & 0    \\
\frac{1}{ {2}}   &  \frac{1}{ {2}}    & 0    \\
0  &  0   & 1
\end{bmatrix}
.
\end{split}
\label{2019-k-e-fireflyrsm-beq}
\end{equation}

\subsubsection{Different intrinsically operational state preparation}

A different approach to partition logic would be to insist that only
{\em intrinsical} -- that is, for any embedded observer having access to means and methods available ``from within'' the system
--  operational state preparations should be allowed.
In such a scenario it is operationally impossible for an observer with access to only one context
-- in the generalized urn model only one color --
to single out the particular type of two-valued measure (aka ball).
Thereby effectively any state preparation is reduced to the
elements of the partition in the respective context (aka color).

Therefore, in the earlier firefly model depicted in Fig.~\ref{2019-k-f-ffl},
the intrinsic operational resolution is among the {\em subsets resulting from the unions of two-valued states} in
$\{1,2\}$,
$\{3,4\}$, and
$\{5\}$ in the first context (aka color);
and among
$\{1,3\}$,
$\{2,4\}$, and
$\{5\}$
in the second context (aka color), as opposed to the single two-valued state discussed earlier in.
Stated differently, an observer accessing a generalized urn
in the first context (aka color) is not capable
to differentiate between the first and the second
two-valued measure (aka ball type),
and would produce a mixture among them if asked to prepare the state
$\{1,2\}$.
Similarly, the observer would not be able to differentiate
between the third and the fourth two-valued measure (aka ball type), and would thus produce a mixture
between those when preparing the state $\{3,4\}$.
However, the ball type $\{5\}$ is recognized and prepared without ambiguity.
Indeed, if one assumes equidistribution (uniform mixtures~\cite[Assumption~1]{Auffeves-Grangier-2015})
of measures (aka ball types),
a very similar situation as in quantum mechanics
[cf Fig.~\ref{2019-k-f-ffl}(d), Eq.~(\ref{2019-k-e-fireflyrsm-beq})]
would result as $\lambda_1=\lambda_2=\lambda_3=\lambda_4=\lambda_5=\frac{1}{5}$
and one would thus ``recover'' the matrix in Eq.~(\ref{2019-k-e-fireflyrsm-beq}).

Pointedly stated there is an epistemic issue of state preparation:
if one demands that the state has to be prepared
by the distinctions accessible from a single context (aka color in the generalized urn model),
then there is no way to prepare or access "ontologic states", say,
selecting balls of type $1$ (first two-valued measure) only.
The difference is subtle:
in the ``ontic'' state case one can resolve (and has access to) every single two-valued measure (aka ball type).
In the ``epistemic,'' intrinsic, operational state case one is limited to the operational procedures available
-- for example, one cannot ``take off the colored glasses''
in Wright's generalized urn model. That is, the resolution of balls
is limited to whatever types can be differentiated in that color.

Whenever such a scenario is considered the respective matrices representing
all conditional probabilities may be very different from the previous scenarios.
Indeed, one may suspect that,
with the assumption of preservation of
equidistributed uniform mixtures across context changes,
the respective matrices are
bistochastic (at least for equidistributed urns) because of a certain type of
``epistemic continuity:'' the sum of the conditional probabilities for
any particular outcome of the second context, relative to all other outcomes
of the first context, should add up to unity.

\begin{figure}[htpb]
\begin{center}
\begin{tikzpicture}  [scale=0.3]

\newdimen\ms
\ms=0.05cm

\tikzstyle{every path}=[line width=1.5pt]

\tikzstyle{c3}=[circle,inner sep={\ms/8},minimum size=4.5*\ms]
\tikzstyle{c2}=[circle,inner sep={\ms/8},minimum size=3*\ms]
\tikzstyle{c1}=[circle,inner sep={\ms/8},minimum size=1.5*\ms]

\newdimen\R
\R=6cm     



\path
  ({90 + 0 * 360 /5}:\R      ) coordinate(1)
  ({90 + 36 + 0 * 360 /5}:{\R * sqrt((25+10*sqrt(5))/(50+10*sqrt(5)))}      ) coordinate(2)
  ({90 + 1 * 360 /5}:\R   ) coordinate(3)
  ({90 + 36 + 1 * 360 /5}:{\R * sqrt((25+10*sqrt(5))/(50+10*sqrt(5)))}   ) coordinate(4)
  ({90 + 2 * 360 /5}:\R  ) coordinate(5)
  ({90 + 36 + 2 * 360 /5}:{\R * sqrt((25+10*sqrt(5))/(50+10*sqrt(5)))}  ) coordinate(6)
  ({90 + 3 * 360 /5}:\R  ) coordinate(7)
  ({90 + 36 + 3 * 360 /5}:{\R * sqrt((25+10*sqrt(5))/(50+10*sqrt(5)))}  ) coordinate(8)
  ({90 + 4 * 360 /5}:\R     ) coordinate(9)
  ({90 + 36 + 4 * 360 /5}:{\R * sqrt((25+10*sqrt(5))/(50+10*sqrt(5)))}     ) coordinate(10)
;


\draw [color=orange] (1) -- (2) -- (3);
\draw [color=red] (3) -- (4) -- (5);
\draw [color=green] (5) -- (6) -- (7);
\draw [color=blue] (7) -- (8) -- (9);
\draw [color=magenta] (9) -- (10) -- (1);    %

%
%
\draw (1) coordinate[c3,fill=orange,label=90:{\footnotesize $\{ 1,2,3\} $}];   %
\draw (1) coordinate[c2,fill=magenta];  %
\draw (2) coordinate[c3,fill=orange,label={above left:\footnotesize $\{ 7,8,9,10,11\}$}];    %
\draw (3) coordinate[c3,fill=red,label={left:\footnotesize $\{ 4,5,6\} $}]; %
\draw (3) coordinate[c2,fill=orange];  %
\draw (4) coordinate[c3,fill=red,label={left:\footnotesize $\{ 1,3,9,10,11\}$}];  %
\draw (5) coordinate[c3,fill=green,label={left:\footnotesize $\{ 2,7,8\} $}];  %
\draw (5) coordinate[c2,fill=red];  %
\draw (6) coordinate[c3,fill=green,label={below:\footnotesize $\{ 1,4,6,10,11\} $}];
\draw (7) coordinate[c3,fill=blue,label={right:\footnotesize $\{ 3,5,9\}$}];  %
\draw (7) coordinate[c2,fill=green];  %
\draw (8) coordinate[c3,fill=blue,label={right:\footnotesize $\{ 1,2,4,7,11\}$}];  %
\draw (9) coordinate[c3,fill=magenta,label={right:\footnotesize $\{ 6,8,10\}$}];
\draw (9) coordinate[c2,fill=blue];  %
\draw (10) coordinate[c3,fill=magenta,label={above right:\footnotesize $\{ 4,5,7,9,11\}$}];  %
\end{tikzpicture}
\end{center}
\caption{Greechie orthogonality diagrams of   the pentagon/pentagram/house logic.
}
\label{2019-k-f-pentagon}
\end{figure}
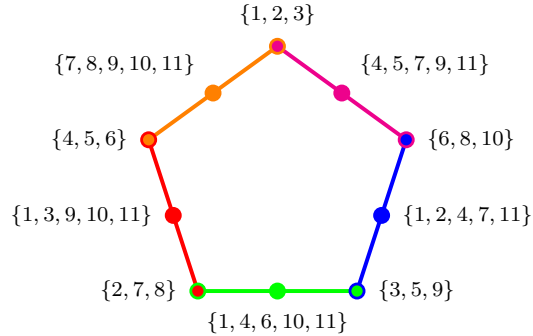

\subsubsection{Pentagon/pentagram/house logic with five cyclically intertwining three-atomic contexts}

By now it should be clear how classical conditional probabilities work on partition logics.
Consider one more example: the pentagon/pentagram/(orthomodular) house~\cite[p.~46, Fig.~4.4]{kalmbach-83}
logic in Fig.~\ref{2019-k-f-pentagon}. Labels of the atoms (aka elementary propositions) are again
obtained by an ``inverse construction''
using all 11 two-valued measures thereon~\cite{wright:pent}.
take, for example, one of the two contexts ${\cal C}_4=\{  \{ 2,7,8\}, \{ 1,3,9,10,11 \}, \{  4,5,6 \} \}$ ``opposite'' to the context
${\cal C}_1=\{  \{ 1,2,3\}, \{ 4,5,7,9,11\}, \{  6,8,10 \} \}$.

With the identifications
${\bf e}_1 \equiv \{1,2,3\}$,
${\bf e}_2 \equiv \{4,5,7,9,11\}$,
${\bf e}_3   \equiv \{6,8,10\}$,
${\bf f}_1 \equiv \{2,7,8\}$,
${\bf f}_2 \equiv \{1,3,9,10,11\}$, and
${\bf f}_3   \equiv \{4,5,6\}$.
The respective conditional probabilities are
\begin{equation}
\begin{split}
\left[ P( {\cal C}_2 \vert {\cal C}_1 )\right] =
\left[ P( \{ {\bf f}_1, {\bf f}_2, {\bf f}_3 \} \vert \{ {\bf e}_1,{\bf e}_2,{\bf e}_3 \} ) \right]
\\
\equiv
\begin{bmatrix}
\frac{P( \{2,7,8\}  \cap  \{1,2,3\} )}{P( \{1,2,3\} )} &  \frac{P( \{1,3,9,10,11\}  \cap  \{1,2,3\} )}{P( \{1,2,3\} )} &  \frac{P( \{4,5,6\}  \cap  \{1,2,3\} )}{P( \{1,2,3\} )}    \\
\frac{P( \{2,7,8\}  \cap  \{4,5,7,9,11\} )}{P( \{4,5,7,9,11\} )}  &  \frac{P( \{1,3,9,10,11\}  \cap  \{4,5,7,9,11\} )}{P( \{4,5,7,9,11\} )}   & \frac{P( \{4,5,6\}  \cap  \{4,5,7,9,11\} )}{P( \{4,5,7,9,11\} )}    \\
\frac{P( \{2,7,8\}  \cap  \{6,8,10\} )}{P( \{6,8,10\} )}  &  \frac{P( \{1,3,9,10,11\}  \cap  \{6,8,10\} )}{P( \{6,8,10\} )}   & \frac{P( \{4,5,6\}  \cap  \{6,8,10\} )}{P( \{6,8,10\} )}
\end{bmatrix}
\\
=
\begin{bmatrix}
\frac{P( \{2\} )}{P( \{1,2,3\} )}  &  \frac{P( \{1,3\} )}{P( \{1,2,3\} )}   & \frac{P( \emptyset)}{P( \{1,2,3\} )}    \\
\frac{P( \{7\} )}{P( \{4,5,7,9,11\} )}  &  \frac{P( \{11\} )}{P( \{4,5,7,9,11\} )}   & \frac{P( \{4,5 \} )}{P( \{4,5,7,9,11\} )}    \\
\frac{P( \{8\} )}{P( \{6,8,10\} )}  &  \frac{P( \{10\} )}{P( \{6,8,10\} )}   & \frac{P( \{6\} )}{P( \{6,8,10\} )}
\end{bmatrix}
\\
=
\begin{bmatrix}
\frac{ \lambda_2 }{ \lambda_1+\lambda_2 +\lambda_3}  &  \frac{ \lambda_1 + \lambda_3 }{\lambda_1+\lambda_2 +\lambda_3 }   &  0    \\
\frac{ \lambda_7 }{ \lambda_4+\lambda_5 + \lambda_7+\lambda_9 +\lambda_{11}  }  &  \frac{ \lambda_9 + \lambda_{11}}{ \lambda_4+\lambda_5 + \lambda_7+\lambda_9 +\lambda_{11}  }   & \frac{ \lambda_4+\lambda_5 }{ \lambda_4+\lambda_5 + \lambda_7+\lambda_9 +\lambda_{11}  }     \\
\frac{ \lambda_8 }{ \lambda_6+\lambda_8+\lambda_{10} }  &  \frac{ \lambda_{10} }{ \lambda_6+\lambda_8+\lambda_{10} }   & \frac{ \lambda_6 }{ \lambda_6+\lambda_8+\lambda_{10} }
\end{bmatrix}.
\end{split}
\end{equation}

\section{Greechie \& Wright's twelfth dispersionless state on the pentagon/pentagram/house logic}

Despite the aforementioned 11 two-valued states there exists another dispersionless state on cyclic pastings of an odd number of contexts; namely, a state being equal to $\frac{1}{2}$
on all intertwines/bi-connections~\cite{greechie-1974,wright:pent}. This state and its associated probability distribution are neither realizable by quantum nor by classical probability distributions.
In this case the conditional probabilities of any two distinct contexts ${\cal C}_i$ and ${\cal C}_j$, for $1\le i,j \le 5$ are
\begin{equation}
\begin{split}
\left[ P( {\cal C}_i \vert {\cal C}_j )\right] \equiv
\begin{bmatrix}
\frac{ 1 }{ 2 }  &  0   &  \frac{ 1 }{ 2 }    \\
0  &  0   &  0  \\
\frac{ 1 }{ 2 }  &  0   &  \frac{ 1 }{ 2 }
\end{bmatrix}
.
\end{split}
\end{equation}

\section{Three-colorable dense points on the sphere}

There exist dense subsets of the unit sphere in three dimensions which require just three colors
for associating different colors within every mutually orthogonal triple of (unit) vectors~\cite{godsil-zaks,meyer:99,havlicek-2000} forming an orthonormal basis.
By identifying two of these colors with the value ``$0$'', and the remaining color
with the value ``$1$'' one obtains a two-valued measure on this ``reduced'' sphere.
The resulting conditional probabilities are discontinuous.

\section{Extrema of conditional probabilities in row and doubly stochastic matrices}

The row stochastic matrices representing conditional probabilities
form a polytope in $\mathbb{R}^{n^2}$ whose vertices
are the $n^n$ matrices $\textsf{\textbf{T}}_i$, $i= 1, \ldots , n^n$, with exactly one entry $1$ in each row~\cite[p.~49]{Berman-Plemmons-NNM}.
Therefore, a row stochastic matrix can be represented as the convex sum $\sum_{i=1}^{n^n} \lambda_i \textsf{\textbf{T}}_i$,
with nonnegative $\lambda_i \ge 0$ and  $\sum_{i=1}^{n^n} \lambda_i =1$.

For conditional probabilities yielding doubly stochastic matrices, such as, for instance, the quantum case, the Birkhoff theorem~\cite{marcus-1960}
yields more restricted linear bounds:
it states that any doubly stochastic $(n\times n)$--matrix is the convex hull of $m\le (n-1)^2+1 \le n!$ permutation matrices.
That is, if $\textsf{\textbf{A}}\equiv a_{ij}$ is a doubly stochastic matrix such that $a_{ij} \ge 0$ and
$
\sum_{i=1}^n a_{ij}
=
\sum_{i=1}^n a_{ji}
=1
$ for $1\le i,j \le n$,
then there exists a convex sum decomposition
$\textsf{\textbf{A}} =\sum_{k=1}^{m \le (n-1)^2+1 \le n!}  \lambda_k \textsf{\textbf{P}}_k$
in terms of $m\le (n-1)^2+1$ linear independent permutation matrices
$\textsf{\textbf{P}}_k$
such that $\lambda_k \ge 0$ and $\sum_{k=1}^{m \le (n-1)^2+1 \le n!}  \lambda_k =1$.

\section{Summary}

I have attempted to sketch a generalized probability theory for multi-context configurations of observables which may or may not be embeddable into a single classical Boolean algebra.
Complementarity and distinct contexts require an extension of the Kolmogorov axioms.
This has been achieved by an additional axiom ascertaining that the conditional probabilities of observables in one context,
given the occurrence of observables in another context, form a stochastic matrix.
Various models have been discussed. In the case of doubly stochastic matrices, linear bounds have been derived from the convex hull of permutation matrices.

\section*{Acknowledgments}
The author acknowledges the support by the Austrian Science Fund (FWF): project I 4579-N and the Czech Science Foundation (GA\v CR): project 20-09869L,
as well as an invitation to the Santiago based IFICC, where enlightening discussions with Tomas Veloz and Philippe Grangier took place.
All misconceptions and errors are mine.


\end{document}